\pdfoutput=1
\documentclass[prx,twocolumn,showpacs,amsmath,amssymb,floatfix,superscriptaddress]{revtex4-1}

\usepackage{graphicx}
\usepackage{dcolumn}
\usepackage{bm}
\usepackage{hyperref}

\begin{document}
	
	\title{Spatio-temporal dynamics of shift current quantum pumping by \\ femtosecond light pulse}
	
	
	\author{U. Bajpai}
	\affiliation{Department of Physics and Astronomy, University of Delaware, Newark, DE 19716-2570, USA}
	\author{B. S. Popescu}
	\altaffiliation{Present Address: Catalan Institute of Nanoscience and Nanotechnology (ICN2), Campus UAB, Bellaterra, 08193 Barcelona, Spain}
	\affiliation{Department of Physics and Astronomy, University of Delaware, Newark, DE 19716-2570, USA}
	\author{P. Plech\'a\v{c}}
	\affiliation{Department of Mathematical Sciences, University of Delaware, Newark,  DE 19716, USA}
	\author{B. K. Nikoli\'{c}}
	\email{bnikolic@udel.edu}
	\affiliation{Department of Physics and Astronomy, University of Delaware, Newark, DE 19716-2570, USA}
	\affiliation{RIKEN Center for Emergent Matter Sciences (CEMS), Wako, Saitama, 351-0198, Japan}
	\author{L. E. F. Foa Torres}
	\affiliation{Departamento de F\'isica, Facultad de Ciencias F\'isicas y Matem\'aticas, Universidad de Chile, Santiago, Chile} 
	\author{H. Ishizuka} 
	\affiliation{Department of Applied Physics, The University of Tokyo, Bunkyo, Tokyo, 113-8656, Japan}
	\author{N. Nagaosa}
	\affiliation{RIKEN Center for Emergent Matter Sciences (CEMS), Wako, Saitama, 351-0198, Japan}
	\affiliation{Department of Applied Physics, The University of Tokyo, Bunkyo, Tokyo, 113-8656, Japan}
	
	
	\begin{abstract}
	Shift current---a photocurrent induced by light irradiating noncentrosymmetric materials in the absence of any bias voltage or built-in electric field---is one of the mechanisms of the so-called bulk photovoltaic effect. It has been traditionally described as a nonlinear optical response of periodic solids to continuous wave light using a perturbative formula,  which is linear in the intensity of light and which involves Berry connection describing the shift in the center of mass position of the Wannier wave function associated with the transition between the valence and conduction bands. Since shift current is solely due to off-diagonal elements of the nonequilibrium density matrix that encode quantum correlations, its peculiar real-space--time dynamics can be expected. We analyze realistic two-terminal devices, where paradigmatic Rice-Mele model is sandwiched between two metallic electrodes, using recently developed time-dependent nonequilibrium Green function algorithms scaling linearly in the number of time steps and capable of treating nonperturbative effects in the amplitude of external time-dependent fields. This unveils novel features: {\em superballistic} transport, signified by time dependence of the displacement, $\sim t^\nu$ with $\nu > 1$,  of the photoexcited charge carriers from the region where the femtosecond light pulse is applied toward the electrodes; and photocurrent quadratic in light intensity at subgap frequencies of light due to two-photon absorption processes that were missed in previous perturbative analyses. Furthermore, frequency dependence of the DC component of the photocurrent reveals shift currents as a realization of nonadiabatic quantum charge pumping enabled by {\em breaking of left-right symmetry} of the device structure. This demonstrates that a much wider class of systems, than the usually considered polar noncentrosymmetric bulk materials, can be exploited to generate nonzero DC component of photocurrent in response to unpolarized light and optimize shift-current-based solar cells and optoelectronic devices.
	
	\bigskip
	
	{\bf KEYWORDS:} Nonlinear optics, ultrafast phenomena, photovoltaics, optoelectronics, nonadiabatic quantum pumping in nanostructures,  time-dependent quantum transport simulations
		
	\end{abstract}
		
	\maketitle


The conventional photovoltaics is based on semiclassical transport of electrons and holes excited by light and separated by the built-in electric field within a $pn$-junction where careful control of disorder and flow of light inside a solar cell is required to reduce entropic energy losses~\cite{Polman2012}. An alternative is distinctly different physical mechanism---the so-called bulk photovoltaic effect (BPVE)~\cite{Sturman1992}---which appears in {\em noncentrosymmetric} systems~\cite{Kral2000a}, such as ferroelectrics with nonzero electric polarization. The BPVE generates steady-state photocurrent and the above band-gap photovoltage as the signatures of nonlinear optical response. However, photocurrent generated by BPVE in a closed circuit remains small compared to $pn$-junctions. This, together with recent understanding of nontrivial topology of electronic bands involved~\cite{Morimoto2016a,Nagaosa2017}, has ignited renewed experimental~\cite{Daranciang2012,Priyadarshi2012,Zenkevich2014,Wachter2015,Spanier2016,Nakamura2017,Ogawa2017} and theoretical~\cite{Young2012,Tan2016,Tan2016a,Cook2017,Rangel2017,Fregoso2017,Ishizuka2017} interest in BPVE with the ultimate goal~\cite{Cook2017} to optimize its power conversion efficiency. BPVE can have contributions from different processes~\cite{Sturman1992,Spanier2016,Baltz1981,Kral2000}, where the most intensely studied are shift current contribution,  traditionally explained as a shift in real space following the carrier interband transition~\cite{Baltz1981,Kral2000,Sipe2000}; and ballistic current contribution due to intraband transition of hot photoelectrons with asymmetric momentum distribution during which they lose their energy to descend to the bottom of conduction band while shifting in real  space~\cite{Sturman1992}. 

The shift current has attracted particular attention due to its inherently quantum-mechanical nature involving coherent evolution of electron and hole wave functions. This makes possible rapid propagation of charge carriers toward the electrodes, thereby minimizing carrier recombination that reduces the magnitude of conventional photocurrent and energy losses due to carrier-phonon scattering~\cite{Kral2000}. For example, very recent experiments~\cite{Nakamura2017,Ogawa2017} shining a laser spot onto the middle of the sample have detected shift current across the whole sample, independently of the position and the width of the excited region, and extracted it \mbox{$\sim 100$ $\mu$m} away from the excited region. Thus, aside from energy harvesting, the shift current photovoltaic effect can also be exploited for sensing and switching applications where the response amplitude and direction depends on both photon energy and polarization while occuring on ultrafast time scales.

The widely used~\cite{Young2012,Tan2016,Tan2016a,Cook2017,Rangel2017,Fregoso2017} ``standard model''~\cite{Kral2000,Sipe2000} for computing the magnitude of shift current is based on a perturbative expansion which yields its DC component, $J_a = \sigma_{abb}(\omega) E_b(\omega)E_b(-\omega)$, as a second-order nonlinear optical response to electric field $E_b(\omega)$ of the incident monochromatic light of frequency $\omega$ and polarization in the $b$-direction [the unpolarized sunlight is modeled by averaging $\sigma_{abb}(\omega)$ over different polarizations]. Thus, it can only describe photocurrent linear in intensity, as also typically encountered in conventional photovoltaics~\cite{Fathpour2007,Ma2014}. The third-rank tensor $\sigma_{abb}(\omega)$ can be expressed as a product of two terms carrying  intuitive  meaning~\cite{Young2012,Fregoso2017,Cook2017}: ({\em i}) diagonal imaginary part of the dielectric function, which is proportional to the density of states (DOS); and ({\em ii}) the so-called shift vector as the average distance traveled by coherent photoexcited electrons during their lifetime. The shift vector depends on the Berry connection~\cite{Morimoto2016a,Nagaosa2017} of the Bloch energy bands of an infinite periodic crystal and it is, therefore, an intrinsic property of the material insensitive to elastic and inelastic scattering processes~\cite{Kral2000}. 

\begin{figure}
	\includegraphics[scale=0.3,angle=0]{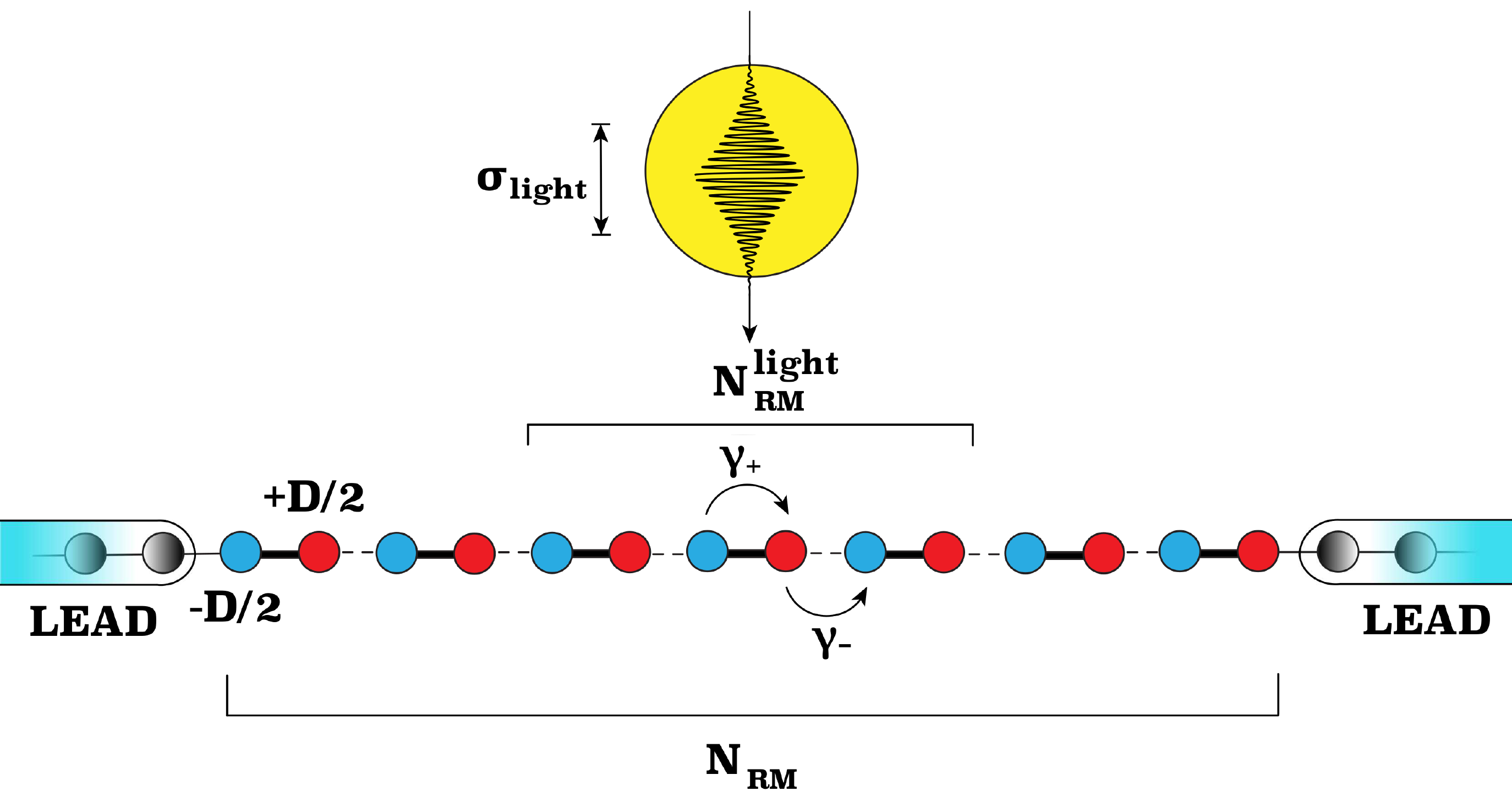}
	\caption{Schematic view of a two-terminal device where photocurrent is generated by a femtosecond light pulse of duration $\sigma_\mathrm{light}$ in the absence of any DC bias voltage between its leads. The central region composed of $N_\mathrm{RM}$ sites is modeled by noncentrosymmetric Rice-Mele TB Hamiltonian in Eq.~\ref{eq:rmh} with alternating hoppings and staggered on-site potential illustrated. It is attached to two-semi infinite NM leads modeled as TB chains with uniform hopping and zero on-site potential. The incident light pulse is assumed to couple to $N_\mathrm{RM}^\mathrm{light} < N_\mathrm{RM}$ sites in the middle of the Rice-Mele central region.}
	\label{fig:fig1}
\end{figure}

Thus, the ``standard model'' tailored for infinite periodic crystals cannot be used to compute open-circuit voltage (as one of  the factors determining the power conversion efficiency~\cite{Tan2016a}) or examine the effect of the electrodes on the measured photocurrent. Furthermore, the ``standard model'' cannot describe photocurrent response to a femtosecond light pulse applied locally to a finite-size open quantum system attached to macroscopic reservoirs, as  illustrated in Fig.~\ref{fig:fig1}. On the other hand, such setups have recently become popular in experiments~\cite{Daranciang2012,Priyadarshi2012,Wachter2015,Nakamura2017,Ogawa2017} aiming to understand: ({\em i}) how fast shift current responds to light and how fast it propagates toward the metallic electrodes~\cite{Nakamura2017,Ogawa2017};  ({\em ii})  how it develops spatially~\cite{Daranciang2012,Priyadarshi2012}; and  ({\em iii}) how it can deviate~\cite{Wachter2015} from the linear dependence on the intensity of light. In this study, we apply time-dependent nonequilibrium Green function (TD-NEGF) formalism~\cite{Stefanucci2013,Gaury2014} to compute, in numerically exact (i.e., nonperturbative) fashion, temporal and spatial development of photocurrent induced by light pulse of duration $\sigma_\mathrm{light}$ and center frequency $\Omega$ irradiating $N_\mathrm{RM}^\mathrm{light}$ sites in the middle of Rice-Mele tight-binding (TB) chain of length $N_\mathrm{RM} \ge N_\mathrm{RM}^\mathrm{light}$.

\begin{figure}
	\includegraphics[scale=0.38,angle=0]{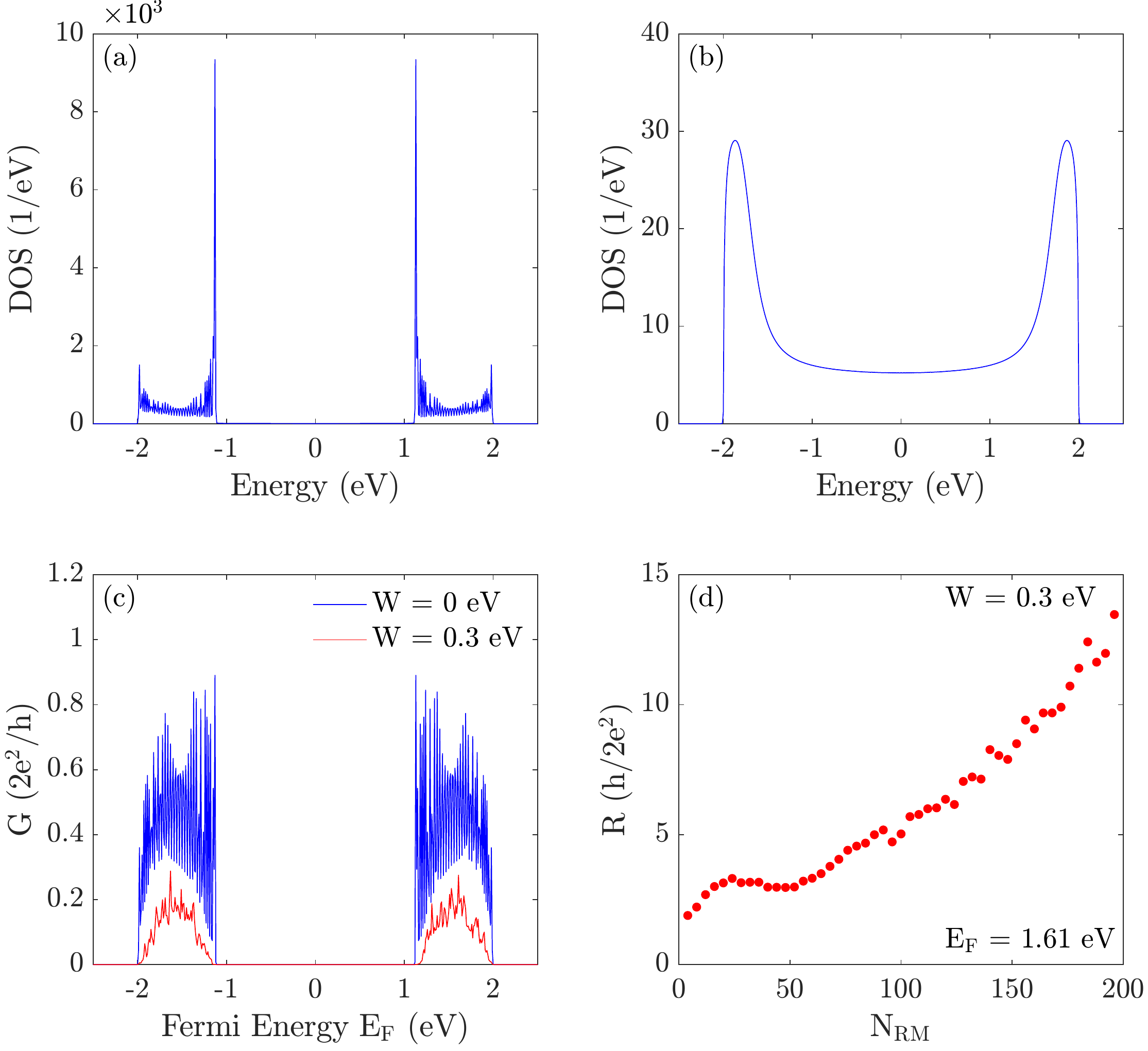}
	\caption{The DOS of the device in Fig.~\ref{fig:fig1} with clean central region of length:  (a) $N_\mathrm{RM}=100$; and (b) $N_\mathrm{RM}=4$ sites. (c) The two-terminal linear-response conductance $G$ of the device in Fig.~\ref{fig:fig1} with $N_\mathrm{RM}=100$ which is clean ($W=0$ eV), or disordered ($W=0.3$ eV) due to a uniform random variable \mbox{$\in [-W/2,W/2]$} added onto  the staggered on-site potential in Eq.~\eqref{eq:rmh}. Panel (d) shows scaling of the two-terminal resistance $R=1/G$ with $N_\mathrm{RM}$ at the Fermi energy \mbox{$E_F=1.61$ eV} and for the same disorder strength \mbox{$W=0.3$ eV} used in panel (c). The geometric disorder averaging to obtain typical $G$ in panel (c) is performed over 25 configurations, and for typical $R$ in panel (d) it is performed over $10^3$ configurations.}
	\label{fig:fig2}
\end{figure}

{\bf Models and Methods.} The Rice-Mele Hamiltonian~\cite{Asboth2016}, as a paradigmatic model employed in shift current studies~\cite{Tan2016a,Ishizuka2017,Rangel2017,Fregoso2017}, is a one-dimensional (1D) model of ferroelectricity along the polar axis
\begin{eqnarray}\label{eq:rmh}
\hat{H}_\mathrm{RM} & = &  \sum_i \left(-\gamma-\frac{B}{2} (-1)^i \right) (\hat{c}^\dagger_{i+1}\hat{c}_i + \hat{c}^\dagger_{i}\hat{c}_{i+1}) \\ \nonumber
&&  + \frac{D}{2} \sum_i (-1)^i \hat{c}^\dagger_{i} \hat{c}_{i},
\end{eqnarray}
whose parameters have been tuned to capture electronic structure of realistic materials like polyacetylene, BaTiO$_3$ and monochalcogenides~\cite{Rangel2017,Fregoso2017}. Here $\hat{c}_i^\dagger$ ($\hat{c}$) creates (annihilates) electron on site $i$; \mbox{$\gamma_+=-\gamma+B/2$} and \mbox{$\gamma_-=-\gamma-B/2$} are the alternating hoppings between the nearest neighbor sites with $B$ parameterizing the dimerization of the chain; and $\pm D/2$ is the staggered on-site potential. The unit cell of Rice-Mele TB chain of size $2a$ contains two sites, and inversion symmetry is broken by $B \neq 0$ and $D \neq 0$~\cite{Tan2016a}. In order to mimic devices used to extract shift current in recent experiments~\cite{Nakamura2017,Ogawa2017}, we attach Rice-Mele TB chain to two semi-infinite normal metal (NM) leads depicted in Fig.~\ref{fig:fig1}, which are modeled by the same TB Hamiltonian in Eq.~\ref{eq:rmh} but with $B=D=0$. We set \mbox{$\gamma=1$ eV} in both the NM leads and in the Rice-Mele central region where \mbox{$B=D=1$ eV} is chosen. The NM leads terminate in the left (L) and right (R) macroscopic reservoirs whose chemical potentials are identical in the absence of DC bias voltage and chosen as $\mu_L=\mu_R=0$ eV (which is band center of NM leads).

For sufficiently long Rice-Mele TB chain, this device exhibits an energy gap \mbox{$\Delta \approx 2.28$ eV} in the DOS [Fig.~\ref{fig:fig2}(a)]. We also consider gapless device where the gap of short Rice-Mele TB chain is filled with evanescent wave functions injected by NM leads [Fig.~\ref{fig:fig2}(b)]. Adding on-site disorder, modeled as a  uniform random variable $\in [-W/2,W/2]$, reduces the van Hove singularities in the DOS at the gap edges while keeping finite two-terminal conductance $G$ outside of the gap  [Fig.~\ref{fig:fig2}(c)]. That is, the scaling of the corresponding two-terminal resistance $R=1/G$ in Fig.~\ref{fig:fig2}(d) with $N_\mathrm{RM}$ shows that for chosen \mbox{$W=0.3$ eV} and $N_\mathrm{RM}=100$ the device is outside of the Anderson localization regime, thereby allowing us to quantify how fast nonequilibrium charge carriers propagate across diffusive TB chain toward the NM leads. We compute $G$ as a linear response to small DC bias voltage between the NM leads using the zero-temperature Landauer formula $G(E_F)=\frac{2e^2}{h}T(E_F)$, where $T(E_F)$ is the transmission function at the Fermi energy $E_F$.

The light pulse is described by the vector potential $\mathbf{A}(t) = A_\mathrm{max} \exp [ -(t-t_p)^2/(2\sigma_\mathrm{light}^2) ] \sin(\Omega t) \mathbf{e}_x$ with a Gaussian shaped function for a pulse of duration $\sigma_\mathrm{light}$, centered at time $t_p$ and center frequency $\Omega$. Here $\mathbf{e}_x$ is the unit vector  along the direction of TB chain, pointing toward the right NM lead in Fig.~\ref{fig:fig1}, which describes linear polarization of incident light. The corresponding electric field is \mbox{$\mathbf{E}=-\partial \mathbf{A}/\partial t$}, while we neglect the relativistic magnetic field of the laser pulse so that electronic spin degree of freedom maintains its degeneracy and it is excluded from our analysis. The vector potential couples to an electron via the Peierls substitution in  Eq.~\eqref{eq:rmh} $\hat{c}^\dagger_{i}\hat{c}_{i+1} \mapsto \hat{c}^\dagger_{i}\hat{c}_{i+1} \exp \left(i z_\mathrm{max}  \exp[-(t-t_p)^2/(2\sigma_\mathrm{light}^2)] \sin(\Omega t) \right)$, where $z_\mathrm{max}=eaA_\mathrm{max}/\hbar$ is the dimensionless parameter quantifying maximum amplitude of the pulse.

To compute local charges and currents driven by time-dependent terms in the Hamiltonian, we employ TD-NEGF formalism which operates with two fundamental quantities~\cite{Stefanucci2013}---the retarded \mbox{$G^{r}_{ii'}(t,t')=-i \Theta(t-t') \langle \{\hat{c}_{i}(t) , \hat{c}^\dagger_{i'}(t')\}\rangle$} and the lesser \mbox{$G^{<}_{ii'}(t,t')=i \langle \hat{c}^\dagger_{i'}(t') \hat{c}_{i}(t)\rangle$} Green functions (GFs) describing the density of available quantum states and how electrons occupy those states, respectively. For the device in Fig.~\ref{fig:fig1} we solve a matrix integro-differential equation~\cite{Croy2009,Popescu2016} for the time-evolution of one-particle reduced nonequilibrium density matrix, \mbox{${\bm \rho}^\mathrm{neq}(t)=\mathbf{G}^<(t,t)/i$},  
\begin{equation}\label{eq:noneqrho}
i\hbar \frac{d {\bm \rho}^\mathrm{neq}}{dt} = [\mathbf{H}_\mathrm{RM}, {\bm \rho}^\mathrm{neq}] + i \sum_{\alpha=\mathrm{L,R}} ({\bm \Pi}_\alpha(t) + {\bm \Pi}_\alpha^\dagger(t)).
\end{equation}
Here all bold-face quantities denote matrices of size $N_\mathrm{RM} \times N_\mathrm{RM}$ in the vector space defined by the central Rice-Mele region of two-terminal device in Fig.~\ref{fig:fig1}. The diagonal elements of the nonequilibrium density matrix yield the local nonequilibrium charge, 
\begin{equation}\label{eq:charge}
Q_i^\mathrm{neq}(t)=e (\rho^\mathrm{neq}_{ii} -  \rho^\mathrm{eq}_{ii}), 
\end{equation}
where we subtract local charge in equilibrium. The equilibrium density matrix can be obtained either as the asymptotic limit  ${\bm \rho}^\mathrm{neq}(t) \rightarrow {\bm \rho}^\mathrm{eq}$ after transient current dies away (and before the light pulse is applied) in the course of time evolution which couples the NM leads to the central Rice-Mele region, or by evaluating \mbox{${\bm \rho}^\mathrm{eq}= -\frac{1}{\pi}\int dE\, \mathrm{Im} \mathbf{G}^r(E) f(E)$} using the retarded GF in equilibrium and the Fermi function $f(E)$ of macroscopic reservoirs~\cite{Areshkin2010}, where we explicitly confirm that both methods give identical result. The matrix 
\begin{equation}\label{eq:current}
{\bm \Pi}_\alpha(t) = \int_{t_0}^t dt_2\, [\mathbf{G}^>(t,t_2){\bm \Sigma}_\alpha^<(t_2,t) - \mathbf{G}^<(t,t_2){\bm \Sigma}_\alpha^>(t_2,t) ],
\end{equation}
is expressed in terms of the lesser/greater GF and the corresponding lesser/greater self-energies~\cite{Stefanucci2013} ${\bm \Sigma}_\alpha^{>,<}(t_2,t)$ whose numerical construction in order to convert Eq.~\eqref{eq:noneqrho} into a system of ordinary differential equations can be found in Ref.~\cite{Popescu2016}. Equation~\eqref{eq:current} yields current in lead $\alpha$ of the device, \mbox{$I_\alpha(t)=\frac{2e}{\hbar} \mathrm{Tr}\, [{\bm \Pi}_\alpha(t)]$}, where summing this expression with the trace of Eq.~\eqref{eq:noneqrho} leads to a continuity equation expressing local charge conservation in time-dependent situations. The local bond  current~\cite{Nikolic2006} between sites $i$ and $j$ connected by the hopping parameter $\gamma_{ij}(t)$ is computed as 
\begin{equation}\label{eq:bond}
I_{i \rightarrow j}(t) =\frac{e}{i\hbar} \left[\rho^\mathrm{neq}_{ij}(t) \gamma_{ji}(t) - \rho^\mathrm{neq}_{ji}(t)\gamma_{ij}(t) \right].
\end{equation}
The computational complexity of TD-NEGF calculations stems from the memory effect---the entire history must be stored in order to accurately evolve the GFs. For efficient calculations over long times and for large number of simulated sites, we employ newly developed TD-NEGF algorithms~\cite{Croy2009,Popescu2016} which scale {\em linearly}~\cite{Gaury2014} in the number of time steps. 

\begin{figure}
	\includegraphics[scale=0.36,angle=0]{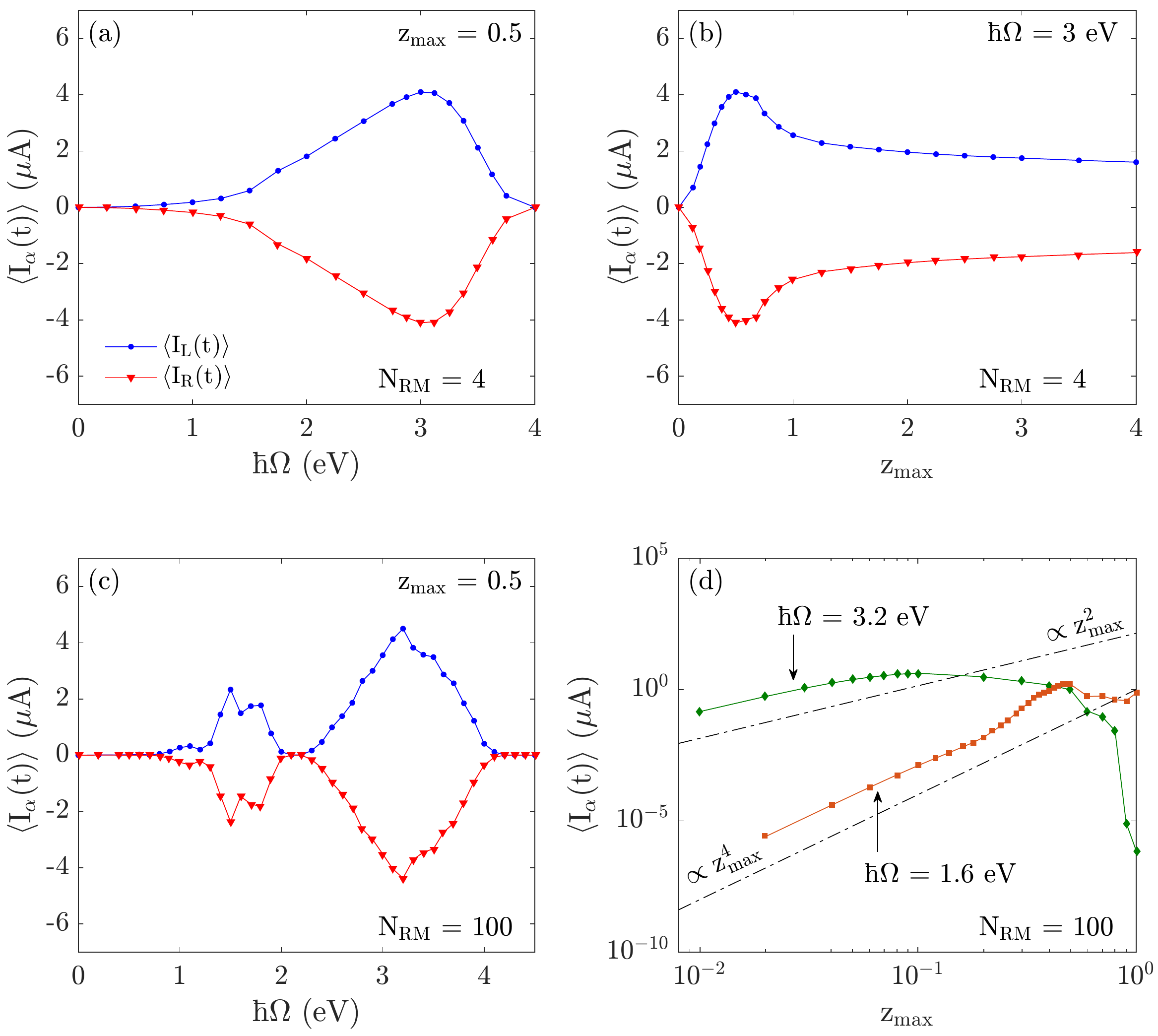}
	\caption{The DC component of photocurrent injected into L and R leads of the clean device in Fig.~\ref{fig:fig1} as a function of (a),(c) center frequency $\Omega$ or (b),(d) maximum amplitude $z_\mathrm{max}$ of light pulse of duration \mbox{$\sigma_\mathrm{light}=50$ fs}.  The pulse irradiates all  $N_\mathrm{RM}=N_\mathrm{RM}^\mathrm{light}=4$ sites of the Rice-Mele TB chain in (a),(b), for which the device in Fig.~\ref{fig:fig1} is gapless with DOS shown in Fig.~\ref{fig:fig2}(b); or it irradiates $N_\mathrm{RM}^\mathrm{light}=20$ middle sites of the Rice-Mele TB chain composed of $N_\mathrm{RM}=100$ sites in (c),(d), for which the device in Fig.~\ref{fig:fig1} is gapped with the DOS shown in Fig.~\ref{fig:fig2}(a).}
	\label{fig:fig3}
\end{figure}
\begin{figure}
	\includegraphics[scale=0.36,angle=0]{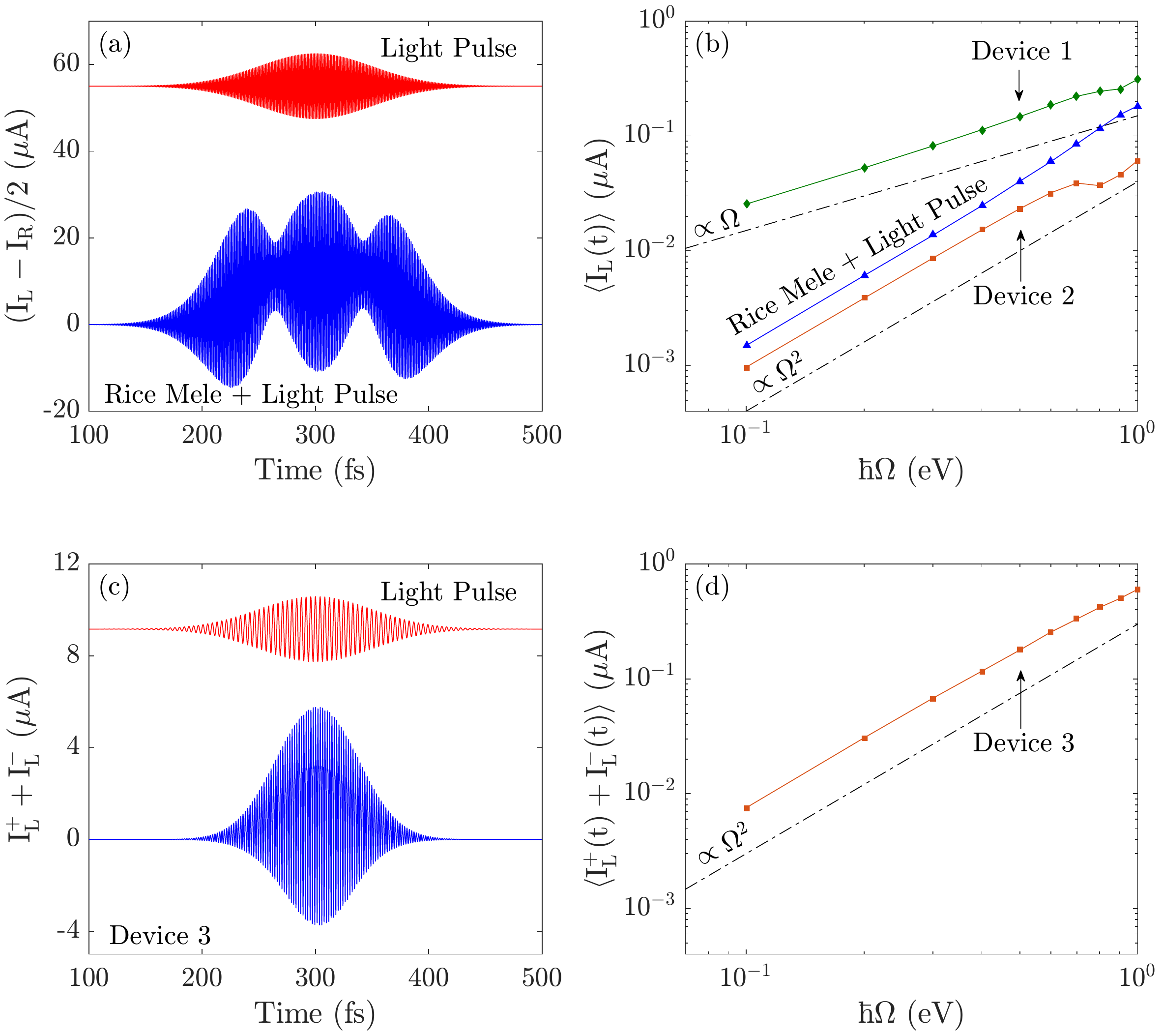} \\ \includegraphics[scale=0.06,angle=0]{fig4e}
	\caption{(a) Time dependence of photocurrent in NM leads attached to Rice-Mele TB chain whose all $N_\mathrm{RM}=N_\mathrm{RM}^\mathrm{light}=4$ sites are irradiated by femtosecond light pulse. (b) Frequency dependence of the DC component of photocurrent in (a), as well as for current pumped into the left NM lead of Rice-Mele TB chain with $N_\mathrm{RM}=4$ sites two of which marked in panel (e) host time-dependent on-site potentials oscillating out-of-phase (Device 1) or in-phase (Device 2) with \mbox{$V_0=1$ eV} and $\Phi=-\pi/3$. (c) Time dependence of total photocurrent in the left NM lead due to unpolarized femtosecond light pulse ($I_\mathrm{L}^{+,-}$ are photocurrents for two different linear polarizations of light) irradiating four sites of an infinite homogeneous TB chain with $B=D=0$ in Eq.~\eqref{eq:rmh}, but including static on-site potentials ($E_1=1$ eV and $E_2=0.5$ eV) at two sites employed to break left-right symmetry of Device 3 in panel (e). Panel (d) shows frequency dependence of the DC component of photocurrent in (c). The light pulse irradiating Rice-Mele TB chain in (a)  has duration \mbox{$\sigma_\mathrm{light}=50$ fs}, maximum amplitude \mbox{$z_\mathrm{max}=0.5$} and center frequency \mbox{$\hbar \Omega = 3$ eV}, while pulse irradiating Device 3 in (c) differs only in the center frequency \mbox{$\hbar \Omega = 1$ eV}.} 
	\label{fig:fig4}
\end{figure}
\begin{figure*}
	\includegraphics[scale=0.6,angle=0]{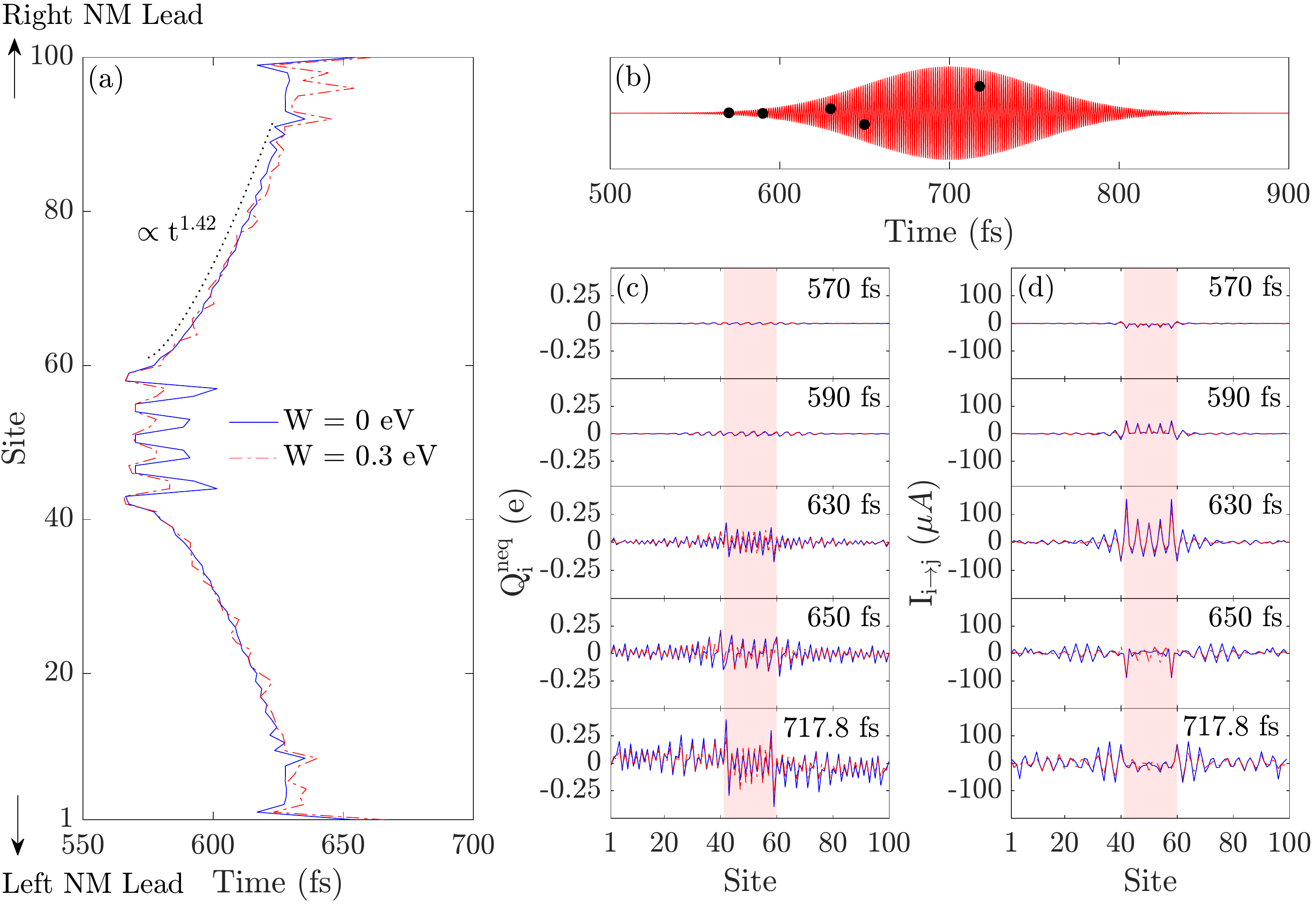}
	\caption{(a) Spatio-temporal profile of nonequilibrium charge $Q_i^\mathrm{neq}(t)$ in Eq.~\eqref{eq:charge} induced by femtosecond light pulse shown in panel (b)---of duration \mbox{$\sigma_\mathrm{light}=50$ fs}, center frequency \mbox{$\hbar \Omega = 3.2$ eV} and maximum amplitude $z_\mathrm{max}=0.5$---irradiating $N_\mathrm{RM}^\mathrm{light}=20$ sites in the middle of a gapped device in Fig.~\ref{fig:fig1} composed of $N_\mathrm{RM}=100$ sites. The curves in (a) trace sites where $|Q_i^\mathrm{neq}(t)|$ at time $t$ reaches 5\% of the maximum value induced [at \mbox{$t=717.8$ fs} in panel (c)] within the irradiated region. Spatial profile at selected times, marked by black dots in panel (b), for: (c) $Q_i^\mathrm{neq}(t)$; and (d) local bond photocurrent $I_{i \rightarrow j}(t)$ between sites of Rice-Mele TB  chain. In panels (a), (c) and (d), solid lines are for clean ($W=0$ eV) and dash-dot lines are for disordered ($W=0.3$ eV) Rice-Mele TB chain. Dotted line in panel (a) represents a power-law fitting of the solid curve in the same panel. The two movies depicting evolution of $I_{i \rightarrow j}(t)$ at all times are provided as Supplemental Material~\cite{sm}.}
	\label{fig:fig5}
\end{figure*}

{\bf Results.} Although electrons of any system will respond  to light pulse by generating a time-dependent photocurrent, photovoltaic applications and the analysis of consequences of broken symmetry are focused on the existence of nonzero DC component of the photocurrent. We define DC component as 
\begin{equation}\label{eq:dc}
\langle I (t) \rangle=\frac{1}{\sigma_\mathrm{curr}}\int I(t) dt, 
\end{equation}
where $\sigma_\mathrm{curr} > \sigma_\mathrm{light}$ is the duration of transient photocurrent, and plot it as a function of $\Omega$ in Fig.~\ref{fig:fig3}(a)  for gapless and in Fig.~\ref{fig:fig3}(c) for gapped device. In gapped device nonzero $\langle I_\alpha(t) \rangle \neq 0$ appears initially at subgap frequency $\hbar \Omega \simeq \Delta/2$, which is at first sight surprising since in the ``standard model'' analyses~\cite{Tan2016a,Fregoso2017} of an infinite Rice-Mele TB chain DC photocurrent is zero in the gap. However, it is compatible with an electron from the valence band in Fig.~\ref{fig:fig2}(a) absorbing two-photons at the same time to transition to the conduction band. Such two-photon absorption mechanism is confirmed by demonstrating  in Fig.~\ref{fig:fig3}(d) scaling  $\langle I_\alpha(t) \rangle \propto z_\mathrm{max}^4$ with the forth power of the electric field or, equivalently, with the square of the intensity of the light pulse. We confirm the same result using charge conserving  Floquet-NEGF approach~\cite{Mahfouzi2012} (truncated to $-1,0,+1$ Floquet bands), where the middle of the Rice-Mele TB chain is irradiated by a continuous-wave (CW) light of frequency $\Omega$. For $\hbar \Omega \gtrsim \Delta$, the DC component of photocurrent is quadratic in electric field or linear in the intensity of the light pulse in Fig.~\ref{fig:fig3}(d), $\langle I_\alpha(t) \rangle \propto z_\mathrm{max}^2$, which is the signature of the usual one-photon absorption mechanism. We note that two-photon photovoltaic effect, as the nonlinear analog of conventional one-photon photovoltaic effect, is rarely observed in standard $pn$-junctions solar cells~\cite{Fathpour2007,Ma2014}.

\begin{figure*}
	\includegraphics[scale=0.36,angle=0]{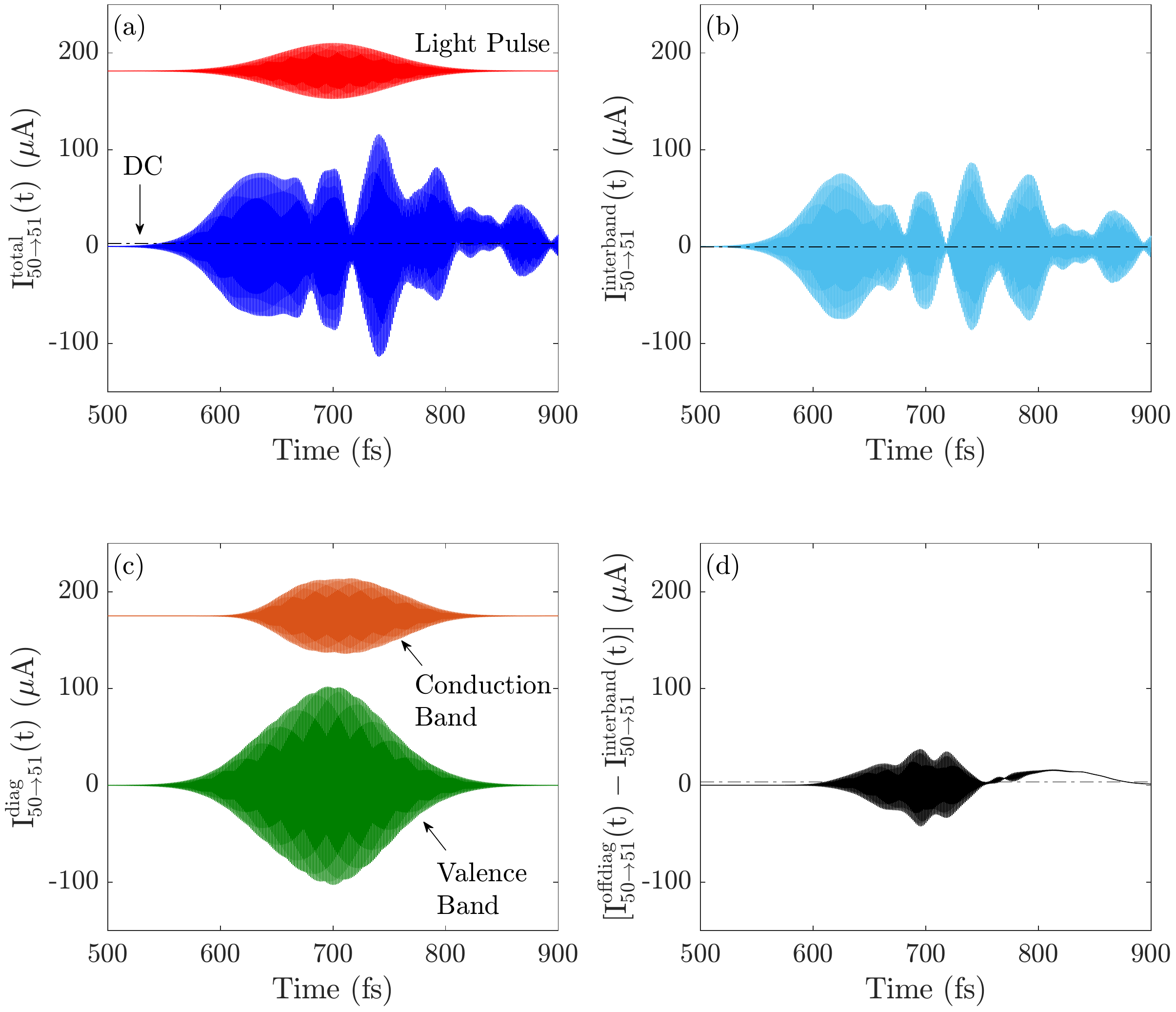} \hspace{0.05in} \includegraphics[scale=0.36,angle=0]{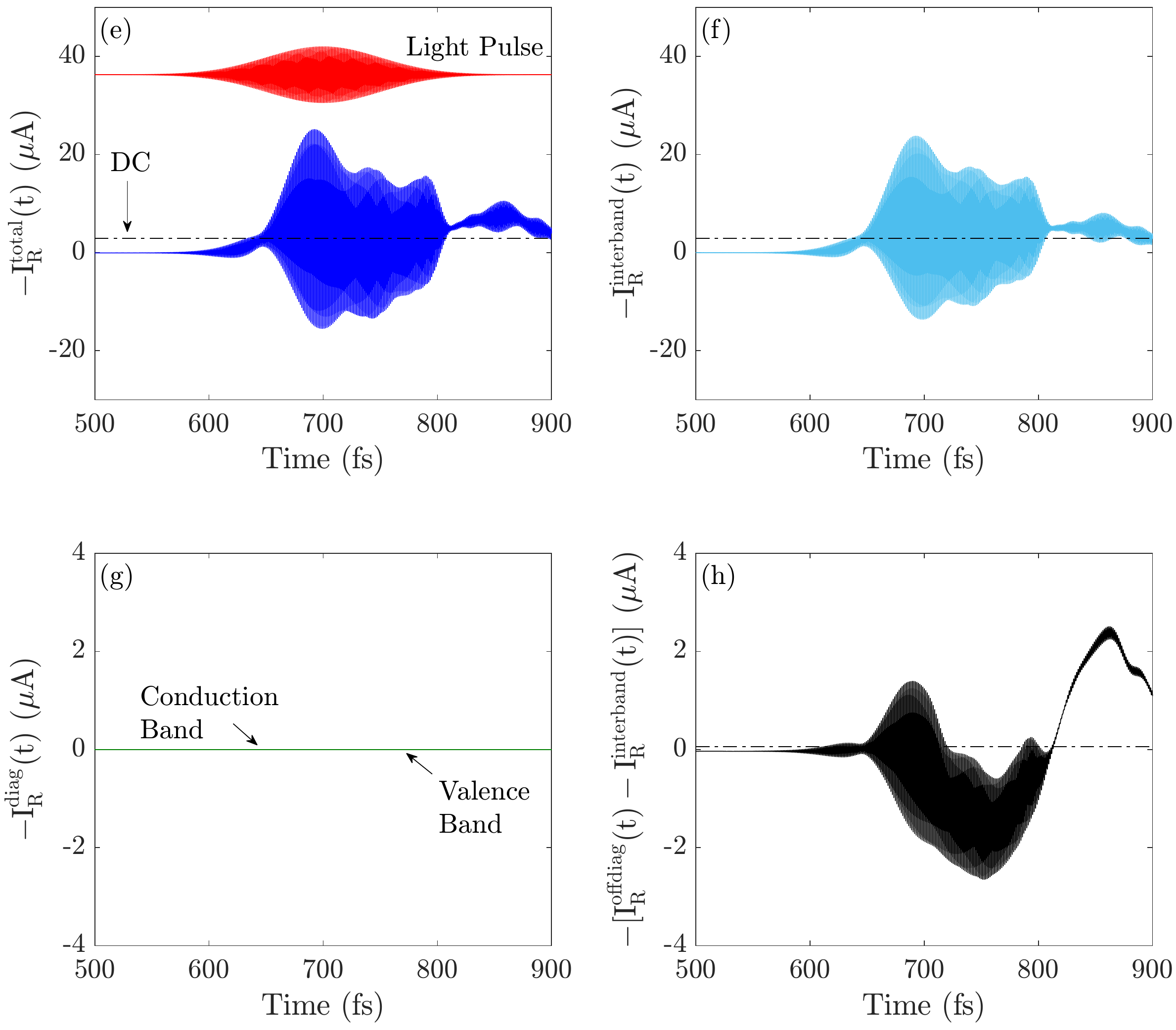}
	\caption{Time dependence of: (a) local bond current $I_{50 \rightarrow 51}(t)$ from site 50 to 51 located in the center of the irradiated region; (b) contribution to $I_{50 \rightarrow 51}(t)$ generated by the  off-diagonal interband elements of nonequilibrium density matrix in Eq.~\eqref{eq:bondenergybasis}, $\langle E_n| \hat{\rho}^\mathrm{neq}  |E_m \rangle$ with $E_n \neq E_m$ belonging to different bands; (c)  contribution generated by the diagonal  elements, $\langle E_n| \hat{\rho}^\mathrm{neq}  |E_n \rangle$ (the conduction band contribution is vertically translated by some constant for clarity);  and (d) contribution generated by the off-diagonal elements, $\langle E_n| \hat{\rho}^\mathrm{neq}  |E_m \rangle$ with $E_n \neq E_m$ belonging to the same band. Panels (e)--(h) plot identical information for bond current flowing from the last site of the Rice-Mele TB chain toward the first site of the right NM lead, which is up to a sign identical to current $I_\mathrm{R}(t)$ collected by the right NM lead [in our convention, positive local bond current flowing into the right NM lead is equivalent to negative $I_\mathrm{R}(t)$].  The dash-dot line in panels (e), (f), (h) denotes the respective DC component defined in Eq.~\eqref{eq:dc}, which is collected by the right NM lead. The pulse---of duration  \mbox{$\sigma_\mathrm{light}=50$ fs}, center frequency \mbox{$\hbar \Omega = 3.2$ eV} and maximum amplitude \mbox{$z_\mathrm{max}=0.5$}---irradiates $N_\mathrm{RM}^\mathrm{light}=20$ sites of Rice-Mele gapped device in Fig.~\ref{fig:fig1} of length $N_\mathrm{RM}=100$ sites.}
	\label{fig:fig6}
\end{figure*}

Figure~\ref{fig:fig4}(a) plots time-dependence of the mean current $[I_\mathrm{L}(t)-I_\mathrm{R}(t)]/2$ injected into the NM leads for gapless device with 
$N_\mathrm{RM}=4$ sites in response to light pulse irradiating all four sites. Its DC component being nonzero in Fig.~\ref{fig:fig3}(a),(b) is in agreement with the key  requirement~\cite{Vavilov2001,Moskalets2002,FoaTorres2005}---{\em breaking of left-right symmetry}---for quantum charge pumping by a time-dependent potential. This, in turn, means breaking inversion symmetry and/or time-reversal symmetry.  In the adiabatic (low frequency) regime, quantum charge pumping requires both inversion and time-reversal symmetries to be broken dynamically, such as by two spatially separated potentials oscillating out-of-phase~\cite{Moskalets2002}, which leads to $\langle I_\alpha(t) \rangle \propto \Omega$ at low frequencies. This is confirmed in Fig.~\ref{fig:fig4}(b) for Rice-Mele TB chain driven by two on-site potentials oscillating out-of-phase [illustrated as Device 1 in Fig.~\ref{fig:fig4}(e)]. In contrast, in the nonadiabatic regime~\cite{Vavilov2001,Moskalets2002,FoaTorres2005}, only one of those two symmetries needs to be broken and this does not have to occur dynamically. The DC component of the pumped current in the nonadiabatic regime is~\cite{FoaTorres2005,Chen2009} $\langle I_\alpha(t) \rangle \propto \Omega^2$ at low frequencies, as confirmed in Fig.~\ref{fig:fig4}(b) for Rice-Mele TB chain driven by either light pulse from panel (a) or by two on-site potentials oscillating in-phase [the latter case is illustrated as Device 2 in Fig.~\ref{fig:fig4}(e)]. 

We note that previous theoretical analyses~\cite{Tan2016a} of shift current in bulk materials have concluded that those with broken inversion symmetry but without electric polarization generate DC component of  photocurrent only in response to polarized light, while nonzero polarization is required to generate DC component of photocurrent in response to  unpolarized light (albeit larger polarization does not always imply a larger photocurrent even though both arise from inversion symmetry breaking)~\cite{Tan2016a,Fregoso2017}. However, these arguments do not consider realistic two-terminal device for which the theory of nonadiabatic quantum charge pumping predicts how leads made of different materials, or identical leads and static on-site potentials~\cite{Chen2009}, employed to break the left-right symmetry of the device structure are sufficient to generate nonzero DC component of photocurrent in response to unpolarized light. We confirm the latter possibility in Fig.~\ref{fig:fig4}(c) by using an infinite TB chain with uniform hoppings and zero on-site potential whose four sites in the middle are irradiated by light pulses of different polarizations ($+$ sign denotes polarization along the TB chain toward the right NM lead and $-$ sign denotes polarization in the opposite direction) while its left-right symmetry is broken by introducing two different static on-site potentials $E_1 > E_2$ [illustrated as Device 3 in Fig.~\ref{fig:fig4}(e)]. This generates photocurrent $I_\mathrm{L}^+(t) + I_\mathrm{L}^-(t)$ [Fig.~\ref{fig:fig4}(c)] in response to unpolarized light whose sum acquires nonzero DC component $\propto \Omega^2$ [Fig.~\ref{fig:fig4}(d)]. 

Motivated by very recent experiments~\cite{Nakamura2017,Ogawa2017}, exploring position dependence of photocurrent induced by applying CW light across the device or its temporal waveform induced by femtosecond light pulse, we examine spatial profiles of nonequilibrium charge [Fig.~\ref{fig:fig5}(c)], $Q_i^\mathrm{neq}(t)$ in Eq.~\eqref{eq:charge},  and local bond current [Fig.~\ref{fig:fig5}(d)], $I_{i \rightarrow j}(t)$ in Eq.~\eqref{eq:bond}, at different times selected [Fig.~\ref{fig:fig5}(b)] within the duration of femtosecond light pulse. The movies depicting spatial profile of $I_{i \rightarrow j}(t)$ at all times are provided as Supplemental Material~\cite{sm}. Both $Q_i^\mathrm{neq}(t)$ and $I_{i \rightarrow j}(t)$ profiles show that at the beginning of the pulse ($t=570$ fs in Fig.~\ref{fig:fig5}) they are localized within the irradiated region composed of middle $N_\mathrm{RM}^\mathrm{light}=20$ sites within $N_\mathrm{RM}=100$ sites Rice-Mele TB chain in Fig.~\ref{fig:fig1}. At later times, they  propagate along non-irradiated region and are eventually collected by the NM leads. 

To understand how fast is the spreading of nonequilibrium charge toward the NM leads, we analyze spatio-temporal profiles depicted in Fig.~\ref{fig:fig5}(a) which trace those sites of clean or disordered Rice-Mele TB chain where absolute value $|Q_i^\mathrm{neq}(t)|$ at time $t$ reaches 5\% (other cutoffs can be used without changing the conclusion) of the maximum value generated [at \mbox{$t=717.8$ fs} in panel (c)] within the irradiated region. Thus, upper  and lower curve in Fig.~\ref{fig:fig5}(a) can be viewed as the  displacement of $Q_i^\mathrm{neq}$ toward the right or left NM lead, respectively. The scaling of the displacement with time, $\sim t^\nu$, can be analyzed akin to variance spreading of optical~\cite{Stutzer2013,Levi2012} or quantum~\cite{Zhang2012d} wave packets or classical Brownian particle~\cite{Bouchaud1990}: $\nu=1$ signifies ballistic propagation in uniform lattices; $\nu=0.5$ or  $\nu=0$ signifies diffusion or Anderson localization in disordered lattices, respectively; and subdiffusion ($0<\nu<0.5$) or  superdiffusion ($0.5<\nu<1$) are also possible in some quasiperiodic lattices~\cite{Abe1987}. However, we find $\nu=1.42(11)$ in the case of clean Rice-Mele TB chain and $\nu=1.28(15)$ in the case of the diffusive one, which demonstrates {\em superballistic} spreading of photoexcited nonequilibrium electrons. We note that faster-than-ballistic spreading of optical wave packets, within certain transient time frame, has been observed experimentally in disordered static~\cite{Stutzer2013} and temporally fluctuating~\cite{Levi2012} photonic lattices, and far-from-equilibrium quantum many-electron system studied in Fig.~\ref{fig:fig5} does share some features with the latter case. The superballistic transport of photoexcited nonequilibrium charge carriers unveiled in Fig.~\ref{fig:fig5}, which evolve quantum-coherently and are largely insensitive to scattering off impurities, is remarkably different from photoexcited carriers in conventional $pn$-junction solar cells where they travel toward the leads via drift-diffusive transport requiring to manipulate their mobilities for efficient DC photocurrent extraction.

{\bf Discussion.} Finally, we show how TD-NEGF framework can be employed to precisely identify shift current contribution to the total photocurrent, thereby removing ambiguities that otherwise lead to controversy~\cite{Fridkin2018} when interpreting experiments~\cite{Nakamura2017} in which some part of a two-terminal device is irradiated by light. Experimentally, the ballistic and shift contributions to BPVE can be distinguished by performing Hall effect measurements because ballistic current is sensitive to external magnetic field and shift current is not~\cite{Zenkevich2014,Spanier2016}. Theoretically, the quantum-mechanical nature of shift current is encoded by the off-diagonal elements~\cite{Baltz1981,Kral2000} of ${\bm \rho}^\mathrm{neq}(t)$, in contrast to ballistic current arising from its diagonal elements. This is not directly visible in thus far presented calculations based on Eq.~\ref{eq:bond}, where local bond current of any origin is extracted from the off-diagonal elements of  ${\bm \rho}^\mathrm{neq}$ in real-space representation while the diagonal elements yield local charge in Eq.~\eqref{eq:charge}. Therefore, we switch to a preferred basis~\cite{Schlosshauer2005} for analyzing quantum coherences encoded in the density matrix by transforming ${\bm \rho}^\mathrm{neq}$ to its representation in the complete basis of eigenstates $|E_n\rangle$ of the Rice-Mele Hamiltonian for an isolated (i.e., not attached to the leads) TB chain of finite length, $\hat{H}_\mathrm{RM} |E_n\rangle = E_n |E_n\rangle$. This allows us to rewrite the bond photocurrent, as the expectation value of bond current operator~\cite{Nikolic2006}, in the following form~\cite{Baltz1981}
\begin{eqnarray}\label{eq:bondenergybasis}
I_{i \rightarrow j}(t) & = & \mathrm{Tr}\, [\hat{\rho}^\mathrm{neq}(t) \hat{I}_{i \rightarrow j}] \nonumber \\ 
&= &\sum_n \sum_m \langle E_n| \hat{\rho}^\mathrm{neq}(t)  |E_m \rangle \langle E_m |\hat{I}_{i \rightarrow j}|E_n\rangle,  
\end{eqnarray}
where we gain access to photocurrent contributions $I_{i \rightarrow j}^\mathrm{diag}(t)$, $I_{i \rightarrow j}^\mathrm{offdiag}(t)$ and $I_{i \rightarrow j}^\mathrm{interband}(t)$ determined by the matrix elements $\langle E_n| \hat{\rho}^\mathrm{neq}  |E_m \rangle$ with $E_n=E_m$ belonging to either the valence or the conduction band in Fig.~\ref{fig:fig2}(a); $E_n \neq E_m$; and $E_n \neq E_m$ with $E_n$ and $E_m$ belong to two different bands, respectively. 

Figure~\ref{fig:fig6}(a)--(d) plots local bond current $I_{i \rightarrow j}(t)$, together with these contributions to it, for a bond $i \rightarrow j$ in the center of the irradiated region of the same gapped device setup studied in Fig.~\ref{fig:fig5}.  Another ambiguity---arising from the traditional theoretical analysis notion that shift current is carried only by photoexcited charge carriers within the irradiated region~\cite{Fridkin2018}---can be resolved by looking at contributions to local bond current far away from the irradiated region, as shown in Fig.~\ref{fig:fig6}(e)--(h) for bond  $i \rightarrow j$ from the last site of the Rice-Mele TB chain toward the first site of the right NM lead. This current is, in fact, the photocurrent $I_\mathrm{R}(t)$ injected into the right NM lead.  The waveforms of the diagonal (i.e., ballistic) contributions $I_{i \rightarrow j}^\mathrm{diag}(t)$ from the conduction and valence bands  follow [Fig.~\ref{fig:fig6}(c)]  that of the light pulse (with some delay in the case $E_n$ belonging to the conduction band) and, therefore, have {\em zero} DC component. Furthermore, such diagonal contributions exist only in the irradiated region [Fig.~\ref{fig:fig6}(c)], while they are identically zero outside of it [Fig.~\ref{fig:fig6}(g)]. 

Thus, the nonzero DC component arises {\em solely} from the off-diagonal contribution $I_{i \rightarrow j}^\mathrm{offdiag}(t)$, which we split into the off-diagonal {\em interband} contribution $I_{i \rightarrow j}^\mathrm{interband}(t)$ [Figs.~\ref{fig:fig6}(b) and \ref{fig:fig6}(f)]  and the off-diagonal {\em intraband} contribution $I_{i \rightarrow j}^\mathrm{offdiag}(t) - I_{i \rightarrow j}^\mathrm{interband}(t)$ [Figs.~\ref{fig:fig6}(d) and \ref{fig:fig6}(h)]. These off-diagonal contributions,  which include quantum coherences from ${\bm \rho}^\mathrm{neq}$ and can be identified as shift current~\cite{Baltz1981}, do not follow the waveform of the light pulse but instead exhibit interference effects, such as in reflection from  Rice-Mele-central-region/NM-lead boundaries. Interestingly, the off-diagonal intraband contribution is also nonzero [Fig.~\ref{fig:fig6}(h)], but it is much smaller than the dominant off-diagonal interband contribution injected into the right NM lead [Fig.~\ref{fig:fig6}(f)].

{\bf Conclusions.} In conclusion, traditional theoretical analyses of photocurrent in noncentrosymmetric systems have been confined to perturbative treatment of an infinite periodic crystal  homogeneously irradiated by CW light which, therefore, cannot explain recent experiments where femtosecond light pulses are applied locally and propagation of photocurrent is studied away from the irradiated region. The nonperturbative TD-NEGF approach introduced in this study makes it possible to analyze spatio-temporal dynamics of photocurrent in realistic devices attached to external leads and exposed to either CW or pulse light of arbitrary intensity. We predict: subgap photocurrent generated by two-photon absorption mechanism; above the gap photocurrent generated by one-photon absorption mechanism; and its superballistic propagation away from irradiated region in both clean and disordered systems. Importantly, we demonstrate that beyond the traditional target materials (i.e., noncentrosymmetric materials with nonzero polarization vector), a much broader class of systems can be explored to optimize the photocurrent in response to unpolarized light (as relevant for photovoltaic applications).  The key requirement defining this broader class of systems is the same as for nonadiabatic quantum charge pumping---broken left-right symmetry of the device structure. For example, it is sufficient to take a two-dimensional material (even non-polar) which interacts with light strongly~\cite{Xia2014,Chen2017b} and attach it to two electrodes made of different metals in order to enable quantum-coherent shift current generation. 


\begin{acknowledgments}
	U. B., B. S. P. and B. K. N. were supported by NSF Grant No. CHE 1566074. P.~P. was supported by ARO MURI Award No. W911NF-14-0247. L. E. F. F. T.  was supported by FondeCyT (Chile) Grant No. 1170917.  H. I. and N. N. were  supported by Japan Society for the Promotion of Science KAKENHI (Grants No. JP16H06717 and JP26103006); ImPACT Program of Council for Science, Technology and Innovation (Cabinet office, Government of Japan, Grant No. 888176); and CREST, Japan Science and Technology (Grant No. JPMJCR16F1). 
\end{acknowledgments}




\end{document}